# LEVERAGING CLOUD BASED BIG DATA ANALYTICS IN KNOWLEDGE MANAGEMENT FOR ENHANCED DECISION MAKING IN ORGANIZATIONS

Mohammad Shorfuzzaman

Department of Computer Science, Taif University, Taif, Saudi Arabia

*ABSTRACT*

*In recent past, big data opportunities have gained much momentum to enhance knowledge management in organizations. However, big data due to its various properties like high volume, variety, and velocity can no longer be effectively stored and analyzed with traditional data management techniques to generate values for knowledge development. Hence, new technologies and architectures are required to store and analyze this big data through advanced data analytics and in turn generate vital real-time knowledge for effective decision making by organizations. More specifically, it is necessary to have a single infrastructure which provides common functionality of knowledge management, and flexible enough to handle different types of big data and big data analysis tasks. Cloud computing infrastructures capable of storing and processing large volume of data can be used for efficient big data processing because it minimizes the initial cost for the large-scale computing infrastructure demanded by big data analytics. This paper aims to explore the impact of big data analytics on knowledge management and proposes a cloud-based conceptual framework that can analyze big data in real time to facilitate enhanced decision making intended for competitive advantage. Thus, this framework will pave the way for organizations to explore the relationship between big data analytics and knowledge management which are mostly deemed as two distinct entities.*

## 1. INTRODUCTION

Knowledge management (KM) is increasingly becoming crucial for enhanced decision making in organizations. Hence, organizations are exploring ways to effectively accumulate and deal with the data, information and knowledge that are accessible today. The earlier development of KM was facilitated by the use of the Information and communications technology (ICT) in late eighties. The goal was to manage the increasing amount of data, information and to ensure its usage and flow across the organization. In the next stage, two pioneers of KM [2] proposed the social, cognitive and business aspects of KM and regarded knowledge as an intellectual asset that an organization should construct and monitor. The size of data and information that are handy today is much more than what we could possibly envisage a decade ago. There is a pressing need to investigate such big amount of data and establish its relationship with KM to enhance organizational decision making and acquire competitive advantage.

The emergence of big data as described by the authors in [24] is due to the drastic increase of processing power, the availability of data with high volume, velocity and variety, and the combination of customary data management and open-sourced technologies and commodity hardware. With the explosion of the Internet, mobile devices, and social media, the impact of big data became apparent in numerous areas and industries. The competence of organizations to make use of big data to derive actionable insights has appeared as a vital policy to build competitive advantages.

This big data due to its various properties like high volume, variety, and velocity can no longer be effectively stored and analyzed with traditional data management techniques [4, 5, 6]. New technologies and architectures are required to store and analyze this data and in turn generate vital real-time information for decision making in organizations. This has opened the door for the researchers to focus on big data analytics which are likely to play a radical role in the success of organizations. The challenge is to collect, store, and analyze the enterprise big data at the right speed from sources such as sales, supply chain, research, and customer relations to build the knowledge base for effective decision making of the organizations. Recent studies also revealed the fact that the utilization of big data has augmented notably in decision-making [3] and both public and private organizations are reaping benefits from this emerging technology [1].

Thus, with the increasingly large amount of data, it is necessary to have a single infrastructure which provides common functionality of big data management, and flexible enough to handle different types of big data and big data analysis tasks. Lately, cloud computing has become an attractive and mainstream solution for data storage, processing, and distribution [7]. It provides on-demand and elastic computing and data storage resources without the large initial investments usually required for the deployment of traditional data centers. Thus, the advent of cloud computing meets the need for large-scale computing and storage infrastructures demanded by big data storage and analytics. Our goal in this regard is to introduce an intelligent cloud-based computing framework that can effectively analyze the big data captured from different enterprise data sources and share this data in real time to facilitate knowledge creation and management towards effective decision making in organizations. Hence, the novelty of approach is to present a unified infrastructure for the organizations to benefit from cloud computing where big data analytics is delivered as an on-demand cloud service and IT resources therein can be quickly adjusted to meet changing demands. As such, the paper provides a detailed account of the relationship between big data analytics and knowledge management from the perspective of effective decision making given the opportunities and challenges faced by cloud computing infrastructures.

The rest of the paper is ordered as follows. Section 2 provides background knowledge and related literature. Requirements analysis for KM in the context of big data analytics is provided in Section 3. Big data driven KM framework is presented in Section 4. Section 5 presents a relevant use case study. Finally, conclusions are given in Section 6 with some potential future work.

## 2. BACKGROUND AND LITERATURE REVIEW

Exploring the role of big data analytics and its relationship with knowledge management is of utmost importance. Hence, in this section, we present a detailed account of big data and knowledge management and their key commonalities with regard to effective decision making in organizations.

### 2.1. BIG DATA

In recent past, big data has become a popular IT phenomenon across business organizations. Organizations are putting more emphasis on collecting the enormous amount of data that are generated everyday due to the tremendous use of electronic devices and IT in product development and dissemination. This huge amount of data is often referred to as "Big Data". Literature shows that there are copious indications suggesting the fact that the use of big data processing and its analysis can harness great values for organizations in developing knowledge management [8]. Prior to finding out how this big data is transformed into knowledge, let us comprehend this further.

There is no agreed upon definition of big data available in the literature. However, big data is generally characterized by three entities such as volume, velocity and variety [9]. Volume represents the large amount of data that are collected which generally range from terabytes to exabytes. This is not adequate to express the real meaning of the concept. The big data presents a great variety which extends beyond structured data, including semi-structured or unstructured data of different types which is usually unsuitable for typical relational databases treatment. They are text, audio, digital pictures and videos posted online, posts to social media sites, log files of the users of social networks, search engines, machine generated data from sensor and device networks, cell phone data to name a few. Finally, this big data is produced with great velocity and must be captured and processed quickly (as in the case of making real time decisions). Figure 1 depicts a wide range of data types from various sources that constitute big data patterns. Some researchers also propose a fourth 'V' which stands for veracity [9]. Veracity refers to the accuracy and credibility of data and the method of analyzing that data. Yet other researchers [25, 26] see this as the *value* of the data that are collected. For example, economic value of some data will vary depending on the origin of the data and it is used.

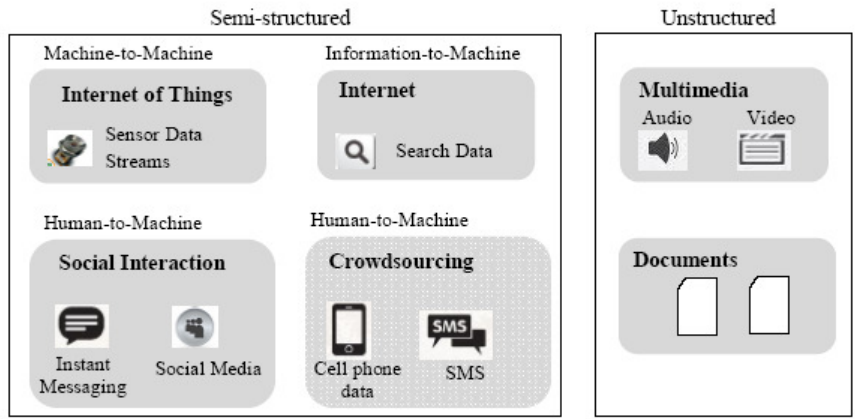

Figure 1. Common sources of big data

Having said the availability of big data in organizations, the rising challenge for them is to process this data to generate useful knowledge to support enhanced decision making. To this end, big data analytics play a significant role in efficient knowledge management that in turns aid in developing business strategic plan in the competitive market and particularly in product development. This entails collection of data from the sources listed above and analyzing the data with big data analytic tools and finally synthesizing the data with relevant enterprise data to obtain constructive knowledge base for storing it and using it further across organizations. To get the utmost benefit of big data, organizations have already initiated the change in their IT infrastructures to deal with the fast rate of extorting and delivering enormous volume of data. More particularly, the organizations with traditional systems would have to comprehend the precedent data and how this can be joined together with their current IT infrastructures.

## 2.2. KNOWLEDGE MANAGEMENT

It is crucial to differentiate between data, information, and knowledge in present-day knowledge literature as shown in Figure 2. Data refers to a set of values either quantitative or qualitative obtained from some events. Information is gathered as a message upon data processing and is communicated generally in a document or audio-visual form. Finally, knowledge has a wider and deeper meaning than data or information. It is created from the use, analysis, and productive utilization of data and information. As shown in the Figure 2, there are two types of knowledge typically covered in knowledge management. They are explicit knowledge and tacit knowledge. Explicit knowledge refers to codified and external knowledge that can be stored in repositories (i.e., as files), intranet, extranet, social media and so on. On the other hand, tacit knowledge exists in the heads of people and in intangible form which is typically communicated and shared through social interactions such as peer-to-peer discussions.

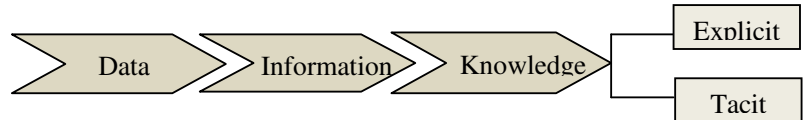

Figure 2. Flow of knowledge creation

The organizations can reap a lot of benefits when this knowledge is managed effectively such as improving customer satisfaction, reducing product development cycle, enhancing product quality and so on. Thus, effective knowledge management plays an important role in decision making and correspondingly guiding the strategic plan for knowledge-based organizations in the context of sustainable competitive advantage [10].

Prior to finding the relationship between big data and knowledge management, let us understand this concept further. Knowledge management is a prescribed structured proposal to advance the creation, distribution or use of knowledge in an organization. This is a formal process of turning corporate knowledge into corporate value. It promotes ongoing business success through the systematic acquisition, synthesis, sharing and use of information insights and experience. More specifically, knowledge management also is the art of creating commercial value from intangible assets. The key business functions of knowledge management is to create, preserve and share knowledge, therefore intellectual capital is important for knowledge management. In a technological sense, knowledge management is the process of acquiring, organizing, sharing and efficiently using organizational knowledge in specific business process, such as production, purchasing, marketing, etc. During sharing and applying, new knowledge is generated and these should be captured and recorded over and over. For instance, in customer service, customer data are captured and stored in appropriate database and then retrieved by the staffs to create service to customers. Thus, new data are generated and then feedback is sent to the system.

Together with a cautious synthesis of knowledge management strategy and implementation of the respective initiatives, an organization can leverage on knowledge management to deliver its value to their customers and to the society. Thriving and sustainable execution of knowledge management can produce many valuable benefits. Firstly, reducing costs and enhancing productivity. Secondly, retaining knowledge from project teams, near retirement staff, and even departed staff. This helps an organization make enhanced consistent decisions. It also facilitates reduced product development or service development cycle. This prevents reinventing the wheel and recommitting any mistakes that were previously made. All these are superb benefits for organizations.

## 2.3. RELATIONSHIP BETWEEN BIG DATA AND KM

We are now in the innovative knowledge ecosystems based on cloud computing and big data. Organizations are looking for means to effectively gather and process the data to create values for improved knowledge management. Recent research has gained much momentum to aid industrial organizations on how to deal with the various organizational and technical prospects and challenges while working with knowledge management, big data, and analytics.

Davenport et al. [10] has outlined a number of potential benefits that organizations can achieve by means of using big data relating to knowledge management. Organizations focus on data flows as opposed to stocks. It is also reported to have increasing reliance on data scientists and product developers rather than data analysts. Finally, they are gradually detaching analytics away from IT tasks and bringing into core businesses and operational functions. In this manner, organizations can create precious knowledge and exploit it for improved knowledge management and competitive market advantage. Thus, it can be inferred that big data and analytics contributes towards real time knowledge management. Big data is also deemed as a knowledge asset and as such state-of-the-art knowledge management has gained substantial impetus due to the use of big data analytics for knowledge creation and management.

To better understand the relationship between big data and KM, we will first look at two approaches to KM: conventional or traditional approach and big data-based approach. Traditional approach typically focuses on conversion of tacit knowledge into explicit knowledge. The normal flow is to capture people's know-how and good practices and then to codify them and put them into repository. On the contrary, big data-based KM deals with discovering the rise of new knowledge based on the huge volume of data that are amassed today. This data is mostly collected from internal sources in addition to external sources, especially from the clouds that we have access to. The focus of big data-based KM will be to do knowledge predictions, knowledge navigation, and knowledge discovery to support enhanced operations and decision making in organizations. These two approaches are in fact not mutually exclusive and can be applied to help businesses and societies at the same time. The vast amount of data that are available in the cloud will change the way companies organize and process business knowledge and provide unparalleled opportunities in knowledge management. This articulates the relationship between big data, knowledge management and cloud computing.

Now, we will closely look at some practical applications of big data in the real world business. Manufacturing companies can use the huge amount of point-of-sale data from all sales outlets and market research data to predict more accurately the fluctuations in demand. This will remove overstock and enable a faster response to out-of-stock situations. Another instance is to make use of sensors to large vehicles such as trucks. Sensors are attached in the engines and tires of a truck fleet to collect data regarding their condition. This collected data are then sent back to a data diagnosis centre for further action. In this manner, the location of the trucks and the wear and tear of important parts are constantly observed in real time before any failure occurs. This can lower downtime costs substantially due to sudden failures and non-delivery of services. Early replacement cost can also be avoided.

The authors in [12] argue that one of the objectives of knowledge management is to assimilate data from different perspectives and analyze them to extract value for effective decision making. This is now a lot simpler to accumulate data from different big data sources and apply big data analytics to generate value from it so that organizations can use it in decision making. For example, expedia.com which is one of the foremost travel websites has invested a large amount of money to use big data analytics to generate valuable insights from the huge amount of data that is generated from everyday use of the site. They analyze the market strategies that attract the customers who visit their website and establish a contributory relationship between their adopted strategies and customers' response. In this manner, the company generates useful insights by analyzing the big data and decides on how to use this valuable information in improving business strategy. This serves as an unambiguous instance of how big data is related to knowledge management. The big challenge faced by industries today to come up with this type of strategic information which help them to make prompt, accurate, and effective tactical decisions.

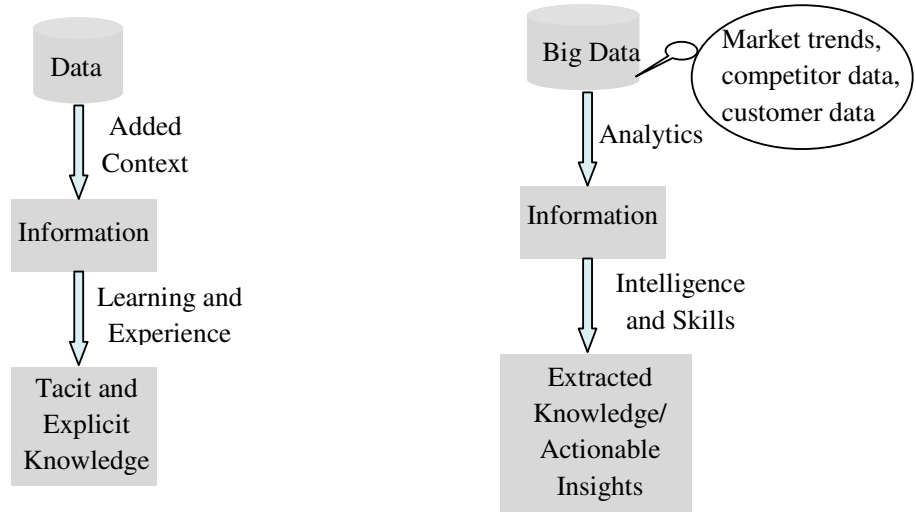

Figure 3. Detailed knowledge creation in traditional and big data based KMS

Credit card companies through big data analytics of huge web monitoring and call center activities data can make better decision regarding personalized customer offers and improved business strategies. In this manner, they are exploiting the concept of real time knowledge generation from big data analytics [11]. Research results obtained by Hair Jr. demonstrate that predictive analytics and big data provide impressive support for emerging business including product development and distribution [13]. The examples given above evidently offer an indication of how big data and knowledge management are interlinked and industries are reaping benefits of big data by generating valuable knowledge. Thus, it can be called as big data based knowledge management that takes full advantage of big data analytics to enhance revenue generation and sales and to reduce risk associated with incurred cost. Figure 3 demonstrates a detailed knowledge creation flow in traditional and big data based knowledge management. Furthermore, Table 1 compares these two types KMS based on a number of factors.

Table 1. Contrast between traditional and big data based knowledge management

| Factor | Traditional Knowledge Management | Big Data based Knowledge Management |
|---|---|---|
| Knowledge type | Both tacit and codified knowledge | Data is processed in real time to extract useful knowledge |
| Necessary skills | Apparently not necessary | Analytical skills are highly necessary |
| Orientation | Highly people oriented | Depends less on people more on machine |
| Interaction | Requires frequent face to face interaction with people | Requires minimal face to face interaction with people |
| Knowledge creation and storage | Tacit knowledge repository is mostly people' brain. Tangible storage of huge size. | Cloud storage is mostly used. Knowledge is created through perpetual flow and processing. |

## 3. REQUIREMENTS ANALYSIS FOR KM IN BIG DATA ANALYTICS CONTEXT

Recent developments in IT have given a boost to the usage of big data and analytics. At the same time, data generation and consumption are driven by the factors such as data standards in organizations, interchange formats of data, and databases used by the organizations [14]. This has motivated organizations to analyze and benefit from the huge amount of data that are obtained from web-based sources, sensors, and mobile devices. Accordingly, knowledge management systems in organizations are receiving momentum by the introduction of big data generation and analytics. To facilitate the processing of such enormous amount of both structured and unstructured data highly optimized databases are necessary. Moreover, it is necessary to present this data so that the user can easily comprehend it. Light-weight web oriented architecture can be used to achieve this [15]. This will facilitate the knowledge discovery and will let the end users realize the finishing outcome in terms of visual entities such as annotated graphs. Thus, to receive the full advantage of big data and analytics, the ICT advancement plays a vital role in big data based KMS. Knowledge sharing is also an important requirement as pointed out in the literature [16] through which an enormous amount of data is produced and thereby knowledge economy grows significantly.

Now, coming into the technology side, big data based KMS appreciates the power of cloud computing. Lately, cloud computing has become an attractive and mainstream solution for data storage, processing, and distribution [7]. It provides on-demand and elastic computing and data storage resources without the large initial investments usually required for the deployment of traditional data centers. Thus, the advent of cloud computing meets the need for large-scale computing and storage infrastructures demanded by big data storage and analytics. The goal in this regard is to introduce an intelligent cloud-based computing framework that can effectively analyze the big data captured from diverse sources for creating knowledge and share this data in real time to facilitate the implementation of a big data based KMS which is effective and fully automated. Hence, organizations can benefit from cloud computing where big data analytics is delivered as an on-demand cloud service and IT resources therein can be quickly adjusted to meet changing demands.

Thus, cloud computing puts forward a vast cyberspace in which knowledge can be created, shared, mined, and synthesized and all these can be done at an unprecedented speed. Big data based KMS envisions to exploit these unprecedented opportunities offered by the cloud computing. Thus, the proposed KMS framework addresses the convergent domain of cloud computing and big data, that are at present two of the most prominent and popular ICT paradigms. The adapted KMS framework hence needs changes that are fundamentally related to transferring the traditional knowledge management system to the exploration of big data deployed in the cloud through services in an effort to a support improved decision making. In addition to storing huge amount of data, cloud computing provides lots of tools for operating on such data. In a cloud, such data and systems are highly accessible and interoperable. Hence, the cloud is appearing as a perfect environment for such experiments to be carried out, to discover new knowledge, to analyze data in order to generate new ideas for organizations to act upon.

As mentioned earlier, the proposed big data based KMS exploits big data analytics for knowledge extraction. The author in [17] suggested that a number of factors need to be considered while taking advantage of big data analytics for successful knowledge creation, deployment, and distribution. First, multidisciplinary team should be formed to supervise the overall process. Second, an appropriate metrics needs to be defined for proper evaluation. Third but not the last, perpetual motivation should be there in terms of special incentives for the work force. In addition, the existing employees need to be trained to gain necessary skills to handle big data analytics in the new system.

The author in [18] explains that big data will be of little value if it cannot help in decision making in organizations. Therefore, a number of key factors are discussed to be considered whenever big data is analyzed for enhanced decision making, improved productivity and efficient operation of the organizations. These will ensure big data to have significant values. First, data should be relevant to a specific course of action. Second, data should be of high quality to ensure dependable and steady support for decision making. Third, huge volume of data even though having potential important information for decision making can result in ineffective outcome if analyzed without vigilant analytic approaches. Fourth, the author in [19] suggests data collection based on the problem at hand to solve. This will help determine the type of data to be gathered based on a number of questions the decision makers will ask.

## 4. BIG DATA DRIVEN KM FRAMEWORK

The challenge in knowledge based systems using ICT oriented solutions lies in the assimilation of data from different sources and processing the data to derive valuable information that is delivered through services, consumed by common citizens, governments, and businesses. However, the lack of a common and standardized platform to meet this challenge severely restricts the creation and sharing of knowledge in organizations. Hence, we propose the development of a unified framework based on cloud computing and big data technologies for the collection, integration and analysis as well as the distribution of real-time data that come from disparate sources to support decision making through automated knowledge management system. To this end, a framework for big data based knowledge management which integrates big data analytics, cloud computing, and knowledge management is illustrated in Figure 4.

We begin with an overview of the system architecture from a technological perspective. Then we describe the architectural patterns for exploiting big data in a knowledge management system. We also describe the functional view on big data, more particularly, the big data analytics technique which is the crux of management system and related big data tools.

### 4.1. INFRASTRUCTURE LAYER

The bottom layer (infrastructure) consists of sensor networks, device networks, and data and management interface. The network components are composed of various types of sensors, actuators, software components, and other devices that collect data from (or actuate on) diverse sources as shown in Figure 1. To manage the underlying network infrastructure and to deal with the data generated therein, a generic control plane (interface) consisting of well-defined data and management APIs (Application Programming Interface) are added. In addition to the big data sources shown in Figure 1, we also consider customers' participation as an important source of data for their welfare, business planning, and decision-making which is generally known as "crowdsourcing" [20]. Overall, this big data is produced with great velocity and must be captured and processed quickly (as in the case of real time monitoring).

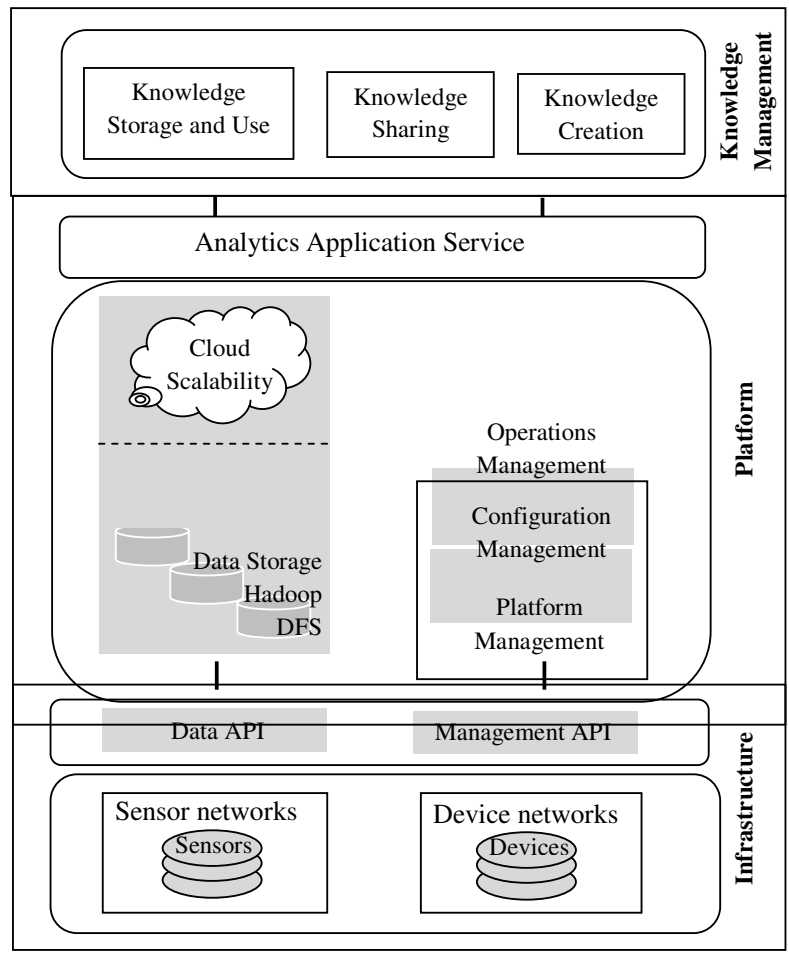

Figure 4. A layered architecture of the framework

### 4.2. PLATFORM LAYER

At the heart of the architecture (as shown in Figure 4) is the second layer (platform) which includes three basic building blocks, namely, big data analytics in the cloud with scalable storage and computing resources, analytics application service, and operations management.

#### 4.2.1. BIG DATA ANALYTICS IN THE CLOUD

The elastic and on-demand nature of resource provisioning makes a cloud platform attractive for the execution of various applications, especially data and computation intensive ones [7]. We plan to integrate cloud computing infrastructure in the proposed framework which includes scalable storage and computation resources to store and analyze the data that are collected from the bottom layer. Thus, big data analytics will be provided as a service by combining it with cloud computing, not as a separate product solution. Instead of deploying self-developed servers, we can actively use the cloud-based big data service to store and analyze the data. At the core of our approach is a new highly scalable big data analytics algorithm that processes various types of data in great volume to discover unknown patterns and relationship and other valuable information for effective decision making in business planning.

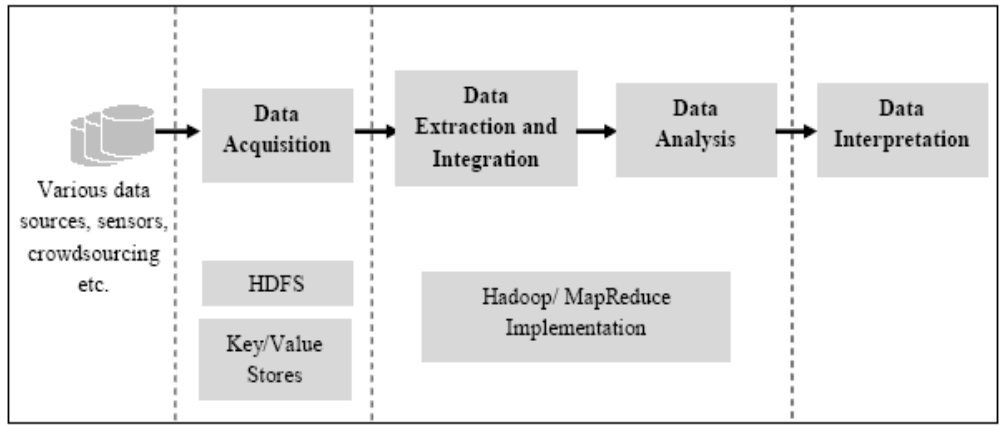

Figure 5. Big data analytics architecture

As shown in Figure 5, the analysis of big data typically involves multiple steps. First, in the acquisition phase, big data are sampled and recorded from various data sources as mentioned above. Second, due to the diversity of the data collected, it's necessary to automatically extract and organize the data for archival, regardless of whether it is real-time data from sensors and other inputs, or data from external systems. Organizing data is also referred to as data integration. Due to the high volume of big data, it is essential to process data in the primary storage location to maintain large throughput and to handle a large variety of data formats, from unstructured to structured. After the above phases, data analysis and modeling will be conducted on the resulting integrated and cleaned big data. The required infrastructure will thus be able to automate decisions based on the analytical model. Finally, data interpretation and visualization will be done since big data analytics alone is of limited value if users cannot understand the analysis results.

#### 4.2.2. REQUIRED TOOLS

The most widely-used technology to handle big data is Hadoop, an open source project based on Google's MapReduce and Google File System [22, 23]. Hadoop is a distributed batch processing infrastructure which consists of the Hadoop kernel, Hadoop Distributed File System (HDFS), and MapReduce programming framework. In our framework, HDFS and/or key-value stores (NoSQL database) will be used to acquire and store big data. They are optimized for exceedingly fast data collection and simple query patterns. In addition, further processing of data will be performed by implementing analytics algorithm using MapReduce framework.

#### 4.2.3. ANALYTICS APPLICATION SERVICE

This service module utilizes the results from the big data analytics for developing domain specific components and services. By this way, users will be able to interact with web-based analytics services easily without worrying about the underlying data storage, management, and analyzing procedures.

#### 4.2.4. OPERATIONS MANAGEMENT

The configuration management module performs configuration of system components and manages network resources. The platform management module provides administrative support for the proposed framework.

### 4.3. KNOWLEDGE MANAGEMENT LAYER

The top layer provides the knowledge management cycle for knowledge acquisition, creation, distribution and sharing, storage, and utilization. Actionable insights are created in real-time to be used in supporting on demand decision making. In particular, real-time analytic results are fed in to the knowledge management layer as input to complete the knowledge management as mentioned. Building on this framework, a community of stakeholders such as industries, government and customers can better plan their business activities and engage in enhanced and effective decision making.

### 4.4 IMPLEMENTATION CHALLENGES

Although the proposed big data based knowledge management framework envisions enjoying the benefits of big data and cloud computing, there are relevant challenges and limitations that need to be tackled. The foremost concern is to deal with technical challenges. Due to the growth of state of the art technologies for collection and managing huge amount of data, the framework needs to find ways to handle large volume of diverse data and time consuming data processing. Often times data is too diverse to analyze and extract value for effective decision making. Another related issue is the poor data governance which acts as an important factor in limiting the efforts of extracting value from big data. At the same time, moving away from traditional data management techniques and coming up with efficient analytical algorithms to process both structured and unstructured data seems to be challenging.

On the other side, the practical implementation of the framework requires the expertise and new tools in cloud based big data technologies. Researchers in the relevant field pointed out that there is a lack of expertise and tools to create value from big data [28]. For instance, big data analytics mostly require the expertise in statistics, machine learning, and data management to extract valuable insight from organizational big data which is going to be a challenge. Hence, successful implementation of the framework will produce new experts in the knowledge management field that exploits these two emerging technologies as such big data and cloud computing.

## 5. A USE CASE STUDY

This section presents a case study conducted by Yahoo big data campaign [27] on predictive modeling for advertising their own products and services. Yahoo represents one of the largest web based companies. The survey statistics show that 80% of the internet users in the United States use Yahoo for diverse internet services. It has experienced over 600 million users from all over the world every month. The services and products offered ranges from miscellaneous content, media, trade and commerce, search and access products. They possess over hundred properties such as mail, television, news, shopping, finance, health, autos, games, movies, travel and so on. The size of the collected data has reached over 25TB every day. This huge amount of data represents the salient source to build content, category, consumer, and campaign actionable insights for their primary content partners and major advertisers. Data processing is conducted in major problem areas such as credit card, travel, stock exchange, retail purchase, and internet usage. Yahoo data challenge thus exceeds other mainstream competitors by at least two orders of magnitude.

Behavioral targeting is one of the important business strategies they adopt to facilitate enhanced decision making intended for competitive advantage. They maintain a behavioral or interest profile and profitability metrics for each consumer. The strategy works by targeting advertisements to customers whose latest behavior or online activities signify which product or service category is related to them. In this manner, the most relevant class of users is identified.

For the sake of completeness, the following diagram illustrates how their predictive modeling works at a very abstract level.

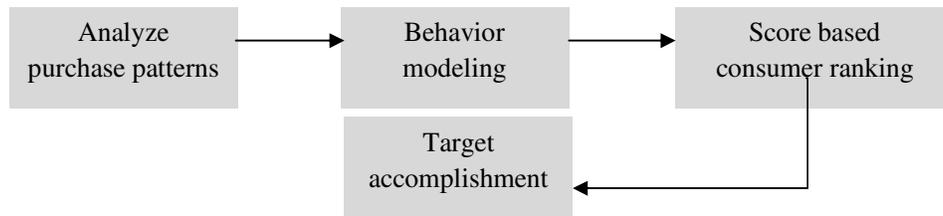

Figure 6. Yahoo predictive modeling cycle for target consumers

The predictive model starts analyzing purchase histories in cycles over hundred of thousands product categories comprising huge volume of data for upfront processing. In each category of these products, a behavior model is built to portray purchase of consumers that presumably represent a huge volume of web click streams containing their responses to ads. In the next stage, each consumer is labeled with a score for appropriate category of product every day. Consumers are then ordered in terms of obtained scores in each category of products and the predictive scheme selects consumers for targeting ads who receive highest relevance scores in the selected category of products.

## 6. CONCLUSIONS

Organizations have realized the importance of huge amount of data that are collected to make enhanced decisions. Research results suggest that organizations with data driven policies are more prolific and profitable compared to their counterparts who are less data driven. Furthermore, evidences demonstrate that big data can create new possibilities and immense opportunities for the organizations to manage knowledge effectively. This paper attempts to investigate the role of big data in knowledge management and make substantial contribution in linking these two concepts. To this end, a cloud based framework is presented to extract the values from big data through analytics for the development of knowledge management to support effective decision making. In the proposed framework, big data platform will use cloud enabling technologies such as HDFS and MapReduce that support distributed processing of data across a cluster of data centers. In addition, NoSQL databases will be used to support real-time processing of big data. The framework exemplifies the complete cycle of big data acquisition and analysis through analytics to transform big data values into actionable insight to support knowledge management process consisting of knowledge acquisition, distribution and sharing, presentation and storage, and utilization. However, a number of challenges are faced in harnessing this new technology in knowledge management such as availability of data scientists and maintaining data privacy and so on which need to be addressed for successful implementation of the proposed framework.